\documentclass{article}

\usepackage{arxiv}
\usepackage{hyperref}
\usepackage{amsmath}
\DeclareMathOperator{\sign}{sign}

\usepackage{amsfonts}

\usepackage{tipa}

\title{Budge: a programming language and a theorem prover}

\author{
  Boro Sitnikovski \\
  Skopje, North Macedonia \\
  \texttt{buritomath@gmail.com} \\
}

\newcommand\doubleplus{\mathbin{{+}\mspace{-8mu}{+}}}

\begin{document}

\maketitle

\begin{abstract}
We present a simple programming language based on Gödel numbering and prime factorization, enhanced with explicit, scoped loops, allowing for easy program composition. Further, we will present a theorem prover that allows expressing and working with formal systems. The theorem prover is simple as it relies merely on a substitution rule and set equality to derive theorems. Finally, we will represent the programming language in the theorem prover. We will show the syntax and semantics of both, and then provide a few example programs and their evaluation.
\end{abstract}

\keywords{Programming language, theorem prover, computational model, Gödel numbering}

\section{Budge programming language}

Budge-PL (b\textturnv\textdyoghlig, b'dzh) is a simple programming language. The programming language uses Gödel numbering\cite{b1} to represent registers and their values by relying on the Fundamental Theorem of Arithmetic\cite{b2}. For example, to represent the values $1$, $2$, and $3$ in memory, we would calculate $2^1 \cdot 3^2 \cdot 5^3$ (the first three primes 2, 3, 5 to the power of the number of the value at the corresponding register), arriving at the state $i = 2250$. We can extract $1$, $2$, and $3$ from $2250$ using prime factorization.

Budge-PL uses similar constructs as FRACTRAN\cite{b3}. However, Budge-PL provides a more convenient way to construct loops and uses integers rather than fractions to denote instructions. A negative integer will decrease a register's value, while a positive integer will increase a register's value. In addition, it provides an easy way to code loops by using nested parenthesis\footnote{One disadvantage of the syntax for loops is that programs can't be as easily represented in memory.}. Finally, it abstracts prime numbers in the code from the programmer.

\subsection{Syntax and semantics}

Where data is represented as $i \in \mathbb{N^+}$ (product of primes), the syntax of the code in Backus-Naur form\cite{b4} is:

\begin{verbatim}
<posn>  ::= "1" | "2" | ...               <negn> ::= "-1" | "-2" | ...
<stmt>  ::= <posn> | <negn> | "("<posn>","<stmts>")"
<stmts> ::= <stmt>","<stmts> | <stmt>     <code> ::= "("<stmts>")"
\end{verbatim}

Let $p(n)$ be the $n$-th prime number. Let $\sign(n) = 1$ if $n>0$ and $-1$ otherwise; this will determine whether we need to multiply or divide. With $\forall x, n_x \in \mathbb{Z} \land n'_x \in \mathbb{Z}$, let $E(i, s)$ represent the evaluation of a sequence $s$ (\texttt{<code>}) for input $i$ be defined by:
$$ E(i, s) = {
\begin{cases}
E(i \cdot p(|n_0|)^{\sign(n_0)}, (n_1, \ldots, n_k)) & s = (n_0, n_1, \ldots, n_k) \land i \cdot p(|n_0|)^{\sign(n_0)} \in \mathbb{N}, \\
E(i, (n_1, \ldots, n_k)) & s = (n_0, n_1, \ldots, n_k) \land i \cdot p(|n_0|)^{\sign(n_0)} \notin \mathbb{N}, \\
E(E(i, (n'_0, \ldots, n'_k)), s) & s = ((P, n'_0, \ldots, n'_k), n_0, \ldots, n_j) \land i \cdot p(|P|)^{-1} \in \mathbb{N}, \\
E(i, (n_0, \ldots, n_j)) & s = ((P, n'_0, \ldots, n'_k), n_0, \ldots, n_j) \land i \cdot p(|P|)^{-1} \notin \mathbb{N}, \\
i & {\text{otherwise, that is, } s = ()}\end{cases}} $$

Semantically, the first case handles increasing/decreasing a value in a register $n_0$. The second case is for skipping an instruction. The third and fourth cases represent the start and end of a loop (nested parenthesis).

\subsection{Example programs}

\subsubsection{Addition of numbers (evaluation explanation)}

To compute $E(2^3 \cdot 3^3, ((2, -2, 1)))$ we iterate the sequence $(-2, 1)$ until $\frac{i}{p(2)}$ is no longer an integer, that is, $\frac{i}{3}$:

\begin{enumerate}
\item Initially, $i = 2^3 \cdot 3^3 = 216$, and since $\frac{216}{3} = 72 \in \mathbb{N}$, proceed with evaluation.
\item Calculate $p(|n|)^{\sign(n)}$ for $n = -2$: $i' = p(2)^{-1} = 3^{-1}$. Since $i \cdot i' \in \mathbb{N}$, set $i$ to $i \cdot i' = 216 \cdot i' = 72$.
\item Calculate $p(|n|)^{\sign(n)}$ for $n = 1$: $i' = p(1)^{1} = 2$. Since $i \cdot i' \in \mathbb{N}$, set $i$ to $i \cdot i' = 72 \cdot i' = 144$.
\item At this point, we go back and check the condition if $\frac{144}{3} \in \mathbb{N}$ - proceed with the evaluation.
\item Calculate $p(|n|)^{\sign(n)}$ for $n = -2$: $i' = p(2)^{-1} = 3^{-1}$. Since $i \cdot i' \in \mathbb{N}$, set $i$ to $i \cdot i' = 144 \cdot i' = 48$.
\item Calculate $p(|n|)^{\sign(n)}$ for $n = 1$: $i' = p(1)^{1} = 2$. Since $i \cdot i' \in \mathbb{N}$, set $i$ to $i \cdot i' = 48 \cdot i' = 96$.
\item At this point, we go back and check the condition if $\frac{96}{3} \in \mathbb{N}$ - proceed with the evaluation.
\item Calculate $p(|n|)^{\sign(n)}$ for $n = -2$: $i' = p(2)^{-1} = 3^{-1}$. Since $i \cdot i' \in \mathbb{N}$, set $i$ to $i \cdot i' = 96 \cdot i' = 32$.
\item Calculate $p(|n|)^{\sign(n)}$ for $n = 1$: $i' = p(1)^{1} = 2$. Since $i \cdot i' \in \mathbb{N}$, set $i$ to $i \cdot i' = 32 \cdot i' = 64$.
\item Now we have that $\frac{64}{3} \notin \mathbb{N}$, so the evaluation halts.
\end{enumerate}

Thus, $i$ is now equal to $64 = 2^6$. That is, the value from the first register $p(1)$ and the value from the second register $p(2)$ were added and then stored in the first register, $p(1)$. In general, $E(2^a \cdot 3^b, ((2, -2, 1))) = 2^n$, with $n = a + b$.

\subsubsection{Other arithmetic operations}

\textit{Subtraction}: $E(2^x \cdot 3^y, s_s) = 2^n \cdot 3^k$ where $n = |x - y|$ and $k = 1$ if $y > x$, and $k = 0$ otherwise.
$$s_s = ((1, -1, 3, 5), (2, -2, 4, 6), (3, -3, -4), (6, -5, -6), (4, -4, 1, 3), (3, (3, -3), 2), (5, -5, 1))$$

\textit{Multiplication}: $E(2^x \cdot 3^y, s_m) = 2^n$ where $n = x \cdot y$.
$$s_m = ((1, -1, (2, -2, 3, 4), (4, -4, 2)), (2, -2), (3, -3, 1))$$

\textit{Division}: $E(2^a \cdot 3^d, s_d) = 2^q \cdot 3^r$ where $a = qd + r$ and $0 \leq r < d$.
\begin{gather*}
s_d = ((2, -2, 7), (1, (7, -7, 2, 8), (8, -8, 7)) \doubleplus s_s \doubleplus \\ (9, (2, -2, (1, -1, -7), (7, -7, 8), -9)), (7, -7), (9, -9, 1), (8, -8, 2))
\end{gather*}

\subsection{Composing and interpreting programs}

As we saw with $s_d$, sequences can be composed by concatenating them: $\forall s_1, \forall s_2, E(E(i, s_1), s_2) = E(i, s_1 \doubleplus s_2)$. For example, the sequence $(1, 2, 2, (2, -2, 1))$ is consisted of concatenating $(1, 2, 2)$ and $((2, -2, 1))$; increasing the first and the second register by 1 and 2 respectively, and then add the registers together, storing the result in the first register.

We show the pseudo-code representation of this sequence by following its semantical interpretation:

\begin{verbatim}
r1 += 1; r2 += 2; // sequence 1
while (r2 > 0) { r2 -= 1; r1 += 1; } // sequence 2
// r1 += r2; r2 = 0; // sequence 2 optimized
\end{verbatim}

\section{Budge theorem prover}

Budge-TP (b\textturnv\textdyoghlig, b'dzh) is a theorem prover that allows expressing formal systems. Formal systems are important because they lie at the core of mathematics. It is directly inspired by Prolog\cite{b5}, where the main difference is that there is no automated deduction and every step has to be manually specified. This allows for a more explicit understanding of formal systems.

Budge-TP has a small Trusted-Computing Base (TCB). Its semantics rely only on substitution and equality check (symbol comparison), though they are still powerful enough to represent any formal system, including computation, as we will see next.

\subsection{Semantics}

Within Budge-TP, a formal system is defined by the tuple $F = (R, V, T)$ together with the functions $subst_{rule}^n$ and $subst_{thm}^n$ where $R_n \in R$ is a set of rules of $n$-ary arguments, $V$ is a set of variables, and $T$ is a set of theorems. A rule $r = (r_1, \ldots, r_n) \in R_n$ is a sequence of string of symbols; it can be roughly interpreted as a function $r_1 \to \ldots \to r_n$, where the $n$-th argument represents a conclusion, and the others represent hypotheses.

Let $S \subseteq V \times T$ denote a set of substitutions, and $X[t/v]$ denote the expression $X$ in which each occurrence of $v$ is replaced with $t$. We define the following function which performs substitution on a rule's hypotheses and conclusion:
$$ subst_{rule}^n(r, S) = {
\begin{cases}
subst_{rule}^n(r_1[t/v], \ldots, r_n[t/v], S \setminus \{(v, t) \}), & r = (r_1, \ldots, r_n) \land (v, t) \in S \\
r & S = \emptyset
\end{cases}}
$$

Let $h = (h_1, \ldots, h_{n-1})$ where $\forall i, h_i \in T$. The function $subst_{thm}^{n-1}(h, S)$ is defined similarly.

For deriving new theorems, we say that $t = subst_{rule}^1((r_n), S) \in T$ (i.e., $t$ is a theorem) if and only if:
$$subst_{rule}^{n-1}((r_1, \ldots, r_{n-1}), S) = subst_{thm}^{n-1}(h, S)$$

Terms and axioms are represented as 1-ary rules; note that for $n = 1$ we have $subst_{rule}^0((), S) = () = subst_{thm}^0((), S)$ i.e. all 1-ary rules are theorems: $\forall r, r \in R_1 \to r \in T$.

\subsection{Syntax}

Even though we used set theory\cite{b6} to represent the semantics, we can liberate from set theory and use a more convenient syntax. Every statement is of the form:

\begin{verbatim}
r<name> : <expr> [-> <expr> [-> ... -> <expr>]]
t<name> : <ruleN> [x=X;y=Y;...] [arg1] [arg2] [...] [argn]
\end{verbatim}

The syntax \texttt{r<name>} specifies a rule, and \texttt{t<name>} specifies a theorem. For \texttt{<name>} and \texttt{<expr>}, any string of characters is accepted except \texttt{':'} and \texttt{' '} (whitespace) for \texttt{<name>} and \texttt{'->'} for \texttt{<expr>}. Square brackets represent optional values. Lowercase characters in a rule expression are considered a variable and will be used for substitution within the expressions.

In a rule, all expressions but the last are considered the hypothesis (arguments to be passed when used in a theorem), and the last is the conclusion. For theorems, the rule \texttt{<ruleN>} will be applied to the corresponding arguments. Substitution with theorems (\texttt{x} with theorem \texttt{X}; \texttt{y} with theorem \texttt{Y}...) will be performed in both the rule's hypotheses and the theorem's provided argument, and they will be matched/unified. If unification is successful, the final argument in the rule \texttt{argn} will be the result.

\subsection{Example theorems}

\subsubsection{MIU system\cite{b1} (set theoretical syntax)}

Let $R = \{ \{ \vdash\texttt{MI}, \texttt{I} \}, \{ (\vdash\texttt{Mx}, \vdash\texttt{Mxx}) \} \}$, $V = \{ \texttt{x} \}$. The particular choice of $R_1$ allows us to pick $S = \{ (\texttt{x}, \texttt{I}) \}$; since \texttt{I} is a 1-ary rule, $\texttt{I} \in T$. Similarly, $\vdash \texttt{MI} \in T$. To prove $\vdash\texttt{MII} \in T$, we use the rule within $R_2$ and since $(\texttt{x}, \texttt{I}) \in S$, we get that $subst_{rule}^1((\vdash\texttt{Mx}), S) = \vdash\texttt{MI} = subst_{thm}^1((\vdash\texttt{MI}), S)$. Since the rule's arguments match the theorem's hypotheses, $subst_{rule}^1((\vdash\texttt{Mxx}), S) = \vdash\texttt{MII} \in T$.

\subsubsection{MIU system (Budge-TP syntax)}

With the following code, we define the terms, the initial axiom and the rules of inference, and a few example theorems:

\begin{minipage}{0.30\textwidth}
\begin{verbatim}
# Terms
rTmM : M
rTmI : I
rTmU : U
tmM! : rTmM
tmI! : rTmI
tmU! : rTmU
rTmxy : xy
\end{verbatim}
\end{minipage}
\begin{minipage}{0.33\textwidth}
\begin{verbatim}
# Axiom and rules
rMI : |- MI
thMI : rMI

r1 : |- xI -> |- xIU
r2 : |- Mx -> |- Mxx
r3 : |- xIIIy -> |- xUy
\end{verbatim}
\end{minipage}
\begin{minipage}{0.33\textwidth}
\begin{verbatim}
# Example theorems
thMII : r2 x=tmI! thMI

tmII! : rTmxy x=tmI!;y=tmI!
thMIIII : r2 x=tmII! thMII

thMUI : r3 x=tmM!;y=tmI! thMIIII
\end{verbatim}
\end{minipage}

The theorems, once deduced, produce the following results:

\begin{verbatim}
thMI : |- MI
thMII : |- MII
thMIIII : |- MIIII
thMUI : |- MUI
\end{verbatim}

\subsubsection{Budge-PL language (Budge-TP syntax)}

Budge-PL uses number theory and prime numbers to store data as registers. We represent a two-register Budge-PL within Budge-TP as a lower system that does not rely on number theory, but rather on a few basic rules:

\begin{minipage}{0.40\textwidth}
\begin{verbatim}
# Lists and numbers
rMkList : (x y)
rTmNil : NIL
rTm0 : 0
rTmS : Sx
rTmP : Px

# Initial program
rInitState : p (a b)
\end{verbatim}
\end{minipage}
\begin{minipage}{0.58\textwidth}
\begin{verbatim}
# Commands 1, -1, 2, -2 respectively
rNextState+1 : (S0 x) (a b) -> x (Sa b)
rNextState-1 : (P0 x) (Sa b) -> x (a b)
rNextState+2 : (SS0 x) (a b) -> x (a Sb)
rNextState-2 : (PP0 x) (a Sb) -> x (a b)

# Commands for looping on the second register
rLoop2Base : ((SS0 x) y) (a 0) -> y (a 0)
rLoop2Succ : ((SS0 x) y) (a Sb)
             -> APPEND x ((SS0 x) y) z -> z (a Sb)
\end{verbatim}
\end{minipage}

Together with the following helper rules for appending lists:

\begin{verbatim}
# Appending lists
rAppendNil : APPEND NIL y y
rAppendRec : APPEND x y z -> APPEND (a x) y (a z)
\end{verbatim}

To calculate the program $((2, -2, 1))$ with the first register 1 and the second 2, we use the rules in order: \texttt{rInitState}, \texttt{rLoop2Succ}, \texttt{rNextState-2}, \texttt{rNextState+1}, \texttt{rLoop2Succ}, \texttt{rNextState-2}, \texttt{rNextState+1}, \texttt{rLoop2Base}. These rules take the initial register state \texttt{(1 2)} and evaluate it to \texttt{(3 0)} which represents the sum.

\end{document}